\documentclass[pra,aps,twocolumn,showpacs,superscriptaddress,floatfix,amsmath,amssymb,nofootinbib]{revtex4}
\usepackage{amsfonts}
\usepackage{graphicx}
\newcommand{\beq}{\begin{equation}}
\newcommand{\eeq}{\end{equation}}

\newcommand{\beqa}{\begin{eqnarray}}
\newcommand{\eeqa}{\end{eqnarray}}
\newcommand{\beqax}{\begin{eqnarray*}}
\newcommand{\eeqax}{\end{eqnarray*}}
\newcommand{\cR}{{\cal{R}}}

\begin{document}
\title{Shortcuts to adiabaticity: fast-forward approach}
\author{E. Torrontegui}
\affiliation{Departamento de Qu\'{\i}mica F\'{\i}sica, Universidad del Pa\'{\i}s Vasco - Euskal Herriko Unibertsitatea, 
Apdo. 644, Bilbao, Spain}

\author{S. Mart\'{\i}nez-Garaot}
\affiliation{Departamento de Qu\'{\i}mica F\'{\i}sica, Universidad del Pa\'{\i}s Vasco - Euskal Herriko Unibertsitatea, 
Apdo. 644, Bilbao, Spain}

\author{A. Ruschhaupt}
\affiliation{Institut f\"ur Theoretische Physik, Leibniz
Universit\"{a}t Hannover, Appelstra\ss e 2, 30167 Hannover,
Germany} 

\author{J. G. Muga}
\affiliation{Departamento de Qu\'{\i}mica F\'{\i}sica, Universidad del Pa\'{\i}s Vasco - Euskal Herriko Unibertsitatea, 
Apdo. 644, Bilbao, Spain}
\begin{abstract}
The ``fast-forward'' approach by Masuda and Nakamura 
generates driving potentials to accelerate slow quantum adiabatic dynamics.
First we present a streamlined version of the formalism that produces the main results in a few steps. Then we show the connection between this approach and inverse engineering based on Lewis-Riesenfeld invariants. We identify in this manner applications in which the engineered potential does not depend on the initial state. Finally we  discuss more general applications exemplified by 
wave splitting processes.   
\end{abstract}  	
\maketitle
\section{Introduction}
Motivated by the practical need to accelerate quantum adiabatic processes in different contexts (transport \cite{David,Calarco,transport, BECtransport,OCTtrans}, expansions \cite{PRL1,OCTexpan}, population inversion and control \cite{Rice03,Rice05,Rice08,Berry2009,PRL2,Morsch}, cooling cycles \cite{Salamon09,PRL1,EPL11}, wavefunction splitting \cite{S07,S09a,S09b,MNProc}), 
and by related fundamental questions 
(about the quantum limits to the speed of processes, 
the viability of adiabatic computing
\cite{Zurek}, or the third principle of thermodynamics \cite{Salamon09,energy}),
a flurry of theoretical and experimental activity   
has been triggered by the proposal of several approaches to design ``shortcuts to adiabaticity''. 
Among other approaches let us mention (i) a transitionless tracking  algorithm or ``counterdiabatic'' approach that adds to the original Hamiltonian 
extra terms to cancel transitions in the adiabatic or superadiabatic bases \cite{Rice03,Rice05,Rice08,Berry2009,PRL2,Morsch}; (ii) inverse engineering of the external driving \cite{GonBEC,PRL1, energy,transport, BECtransport, NHSara, 3d, LabeyriePRA, BECexp} based on Lewis-Riesenfeldt invariants \cite{LR}, which has been applied in several expansion experiments \cite{LabeyriePRA,BECexp}; (iii) optimal control  (OC) methods \cite{S07,Salamon09,OCTexpan,OCTtrans}, sometimes combined with other methods to enhance their performance \cite{OCTexpan, BECtransport, OCTtrans}; (iv) the fast-forward (FF) approach advocated by Masuda and Nakamura \cite{MNProc,MN11}; (v) parallel adiabatic passage \cite{PLAP1,PLAP2,PLAP3,PLAP4}.    
 
The multiplicity of approaches is quite useful because they 
may complement each other: either in the same application,     
as demonstrated e.g. with OC and invariant-based methods, or because of their different 
domains. 
Clarifying the features, overlaps, and relations among these approaches is important to apply the ones which are best suited for specific systems and objectives \cite{CTM}, 
or to develop new ones. 
In this paper we shall establish in particular the connection between the fast-forward and the invariant-based methods. 

Based on some earlier results \cite{MNPra}, the fast-forward formalism for adiabatic dynamics and several application examples were worked out in \cite{MNProc, MN11} by Masuda and Nakamura
for the Gross-Pitaevskii (GP) or the corresponding Schr\"odinger equations. 
The objective of the method is to accelerate a ``standard'' system subjected to a slow variation of external parameters. The time is then rescaled by a ``magnification factor'',
and an ansatz wave function is defined by the standard function multiplied by a
phase factor that, in general, depends on position and time. Inserting the ansatz into the dynamical equation provides in principle the form of the necessary fast-forward driving potential and the equation to be satisfied by the phase. This procedure leads however to divergent terms when the reference standard process is infinitely slow. 
The solution found in \cite{MNProc} to this problem   
was to ``regularize'' the Hamiltonian and standard state using a new ansatz.

As a consequence of the different steps and functions introduced the resulting procedure is somewhat involved, which hinders a broader application.  
In Sec. \ref{secmethod} we provide a streamlined construction of 
local and real fast-forward potentials, and Sec. \ref{secinverse} 
delves into a more detailed connection between this streamlined version and the original formulation of the FF formalism. Section \ref{secconexion} relates the fast-forward approach to the inverse method for dynamical invariants which are quadratic in momentum. Section \ref{beyond} discusses applications beyond this domain, in particular wavefunction splitting, which is an important operation for  
matter wave interferometry \cite{S07,S09a,S09b,Augusto}.    
Finally Sec. \ref{secoutlook}  
discusses the results and open questions.


\section{A simple inverse method\label{secmethod}}
Our starting point is the 3D time-dependent GP equation
\beq
\label{start}
i\hbar\frac{\partial|\psi(t)\rangle}{\partial t}=H(t)|\psi(t)\rangle,
\eeq
where the Hamiltonian $H$ is the sum of the  kinetic energy $T$, the external potential $V(t)$,
and  the mean field potential $G(t)$. We are assuming an external local
potential, where ``local'' means here $\langle \bold{x}|V(t)|\bold{x'}\rangle=V(\bold{x},t)\delta(\bold{x}-\bold{x'})$.
Then, by solving Eq. (\ref{start}) in coordinate space, $V(\bold{x},t)$ 
may be written as 
\beq
\label{pot1}
V(\bold{x},t)=\frac{i\hbar\langle \bold{x}| \partial_t\psi(t)\rangle-\langle\bold{x}|T|\psi(t)\rangle-\langle\bold{x}|G(t)|\psi(t)\rangle}{\langle\bold{x}|\psi(t)\rangle}, 
\eeq 
with $\langle\bold{x}|\psi(t)\rangle=\psi(\bold{x},t)$.
The kinetic and mean field terms in the coordinate representation have the usual forms
\beqax
\langle\bold{x}|T|\psi(t)\rangle&=&\frac{-\hbar^2}{2m} \nabla^2\psi(\bold{x},t),
\\
\langle\bold{x}|G(t)|\psi(t)\rangle&=&g|\psi(\bold{x},t)|^2\psi(\bold{x},t),
\eeqax
$g$ being the coupling constant of the Bose-Einstein condensate. 
By introducing into Eq. (\ref{pot1}) the ansatz
\beq
\label{wave}
\langle\bold{x}|\psi(t)\rangle=r(\bold{x},t)e^{i\phi(\bold{x},t)}, \quad r(\bold{x},t), \phi(\bold{x},t) \in \mathbb{R},
\eeq
we get 
\beqa
V(\bold{x},t)&=&i\hbar\frac{\dot{r}}{r}-\hbar{\dot \phi}+\frac{\hbar^2}{2m}\bigg(\frac{2i\nabla \phi\cdot\nabla r}{r}+i\nabla^2\phi
\nonumber \\
&-&(\nabla \phi)^2+\frac{\nabla^2r}{r}\bigg)-gr^2,
\label{pot2}
\eeqa
where the dot means time derivative. 
The real and imaginary parts are
\beqa
{\rm{Re}}[V(\bold{x},t)]&=&-\hbar{\dot \phi}+\frac{\hbar^2}{2m}\bigg(\frac{\nabla^2 r}{r}-(\nabla \phi)^2\bigg)-gr^2, \label{real}
\\
{\rm{Im}}[V(\bold{x},t)]&=&\hbar\frac{\dot r}{r}+\frac{\hbar^2}{2m}\bigg(\frac{2\nabla \phi\cdot \nabla r}{r}+\nabla^2 \phi\bigg).
\label{imag}
\eeqa
Our purpose now is to design a local and real potential 
such that an initial eigenstate of the initial Hamiltonian, typically the ground state but it could be otherwise, evolves in a time $t_f$ into the corresponding eigenstate of the final
Hamiltonian (a different goal will be discussed in the final Section). 
%
%
%
We assume that 
the full Hamiltonian and the corresponding eigenstates are known at the boundary times.

By construction the potential of Eq. (\ref{pot2}) is local. If we impose ${\rm{Im}}[V(\bold{x},t)]=0$, i.e.  
\beq
\frac{\dot r}{r}+\frac{\hbar}{2m}\bigg(\frac{2\nabla \phi\cdot \nabla r}{r}+\nabla^2 \phi\bigg)=0,
\label{imag0}
\eeq
then we get from Eq. (\ref{real}) a local and real potential.
  
In the inversion protocol we design $r(\bold{x},t)$   
first, 
then solve for $\phi$ in Eq. (\ref{imag0}), and finally get the potential $V$ from Eq. (\ref{real}). 
If, at the boundary times, $\dot{r}=0$ is imposed, Eq. (\ref{imag0}) has
solutions $\phi(\bold{x},t)$ fulfilling that
$\phi(\bold{x},t)$ is independent of $\bold{x}$ at $t=0$ and $t=t_f$.  
Using this in Eq. (\ref{real}) at $t=0$, and multiplying by $e^{i\phi(0)}$, we get  
\beq
\bigg[-\frac{\hbar^2}{2m}\nabla^2+V(\bold x,0)+g|\psi(\bold x,0)|^2\bigg]\psi(\bold x,0)=-\hbar\dot \phi(0)\psi(\bold x,0). 
\eeq 
The initial state is thus an eigenstate of the stationary GP equation at $t=0$,
and $-\hbar\dot\phi(0)=E(0)$ is the energy of
the eigenstate $\psi(\bold x,0)$. Note that 
the above solution of $\phi$ (with $\dot{r}=0$ at boundary times)
admits the addition of an arbitrary function that depends only on time and modifies the zero of energy. 
A similar result is found at $t_f$.


\section{Connection with the fast-forward approach\label{secinverse}}
We shall now reformulate the above results to conect them with the FF approach in  \cite{MNPra, MNProc}. The notation is made close but not necessarily in full agreement with \cite{MNPra, MNProc}.  
Let us define an external parameter that depends on time, or on some scaled time function, according to   
\beq
R(\Lambda(t))=\epsilon\Lambda(t) =: \cR(t).  
\eeq
Here $\cR(t)$ and $R(\Lambda)$ are in general different functions of their arguments,  
$\epsilon$ is a small positive constant and
the scaling function $\Lambda(t)$ is given in terms of a magnification factor $\alpha$,
\beq  
\Lambda(t)=\int_{0}^{t}\ dt'\alpha(t'). 
\label{rt}
\eeq
$\alpha(t)$ is positive 
for $0\le t \le t_f$ and zero at the boundaries $t=0$ and $t=t_f$.
Note that $\dot\cR=\epsilon\alpha$ and $\ddot\cR=\epsilon\dot\alpha$.
We rewrite the modulus and the phase in Eq. (\ref{wave}) as
\beqa
r(\bold x,t)&=& \tilde{r}(\bold x,\cR(t)),
\label{eqr}\\
\phi(\bold x,t)&=&-\frac{1}{\hbar} \int_0^t
dt'{\cal{E}}(\cR(t'))+\epsilon\alpha(t)\theta(\bold x,\cR(t)),\label{eqphi}
\eeqa 
where again we have distinguished the functions according to their different arguments, 
in particular $E(t)={\cal{E}}(\cR(t))$. 
If we also demand $\dot\alpha=0$ at the boundaries to fulfill $\dot r=0$,
then
$\dot{\phi}(\bold x,0)=-{\cal{E}}(\cR(0))/\hbar$ and $\dot{\phi}(\bold x, t_f)=-{\cal{E}}(\cR(t_f))/\hbar$.  
Substituting Eq. (\ref{eqphi}) in Eq. (\ref{imag0}), $\theta$  has to satisfy
\beqa
0&=&\tilde{r}(\bold{x},\cR(t)) \nabla^2\theta(\bold{x},\cR(t))+2\nabla\tilde{r}(\bold{x},\cR(t)) \cdot \nabla\theta(\bold{x},\cR(t))\nonumber\\
&+&\frac{2m}{\hbar}\frac{\partial \tilde{r}}{\partial \cR}(\bold{x},\cR(t)),
\label{Masuda1}
\eeqa
and from Eq. (\ref{real}), the ``fast-forward'' potential is given by
\beqa
V(\bold x,t)&=& V_0(\bold{x},\cR(t))-\hbar\epsilon\dot\alpha(t)\theta(\bold{x},\cR(t))\nonumber\\
& &-\hbar\epsilon^2\alpha^2(t)\frac{d\theta}{d\cR}(\bold{x},\cR(t))\nonumber\\
& &-\frac{\hbar^2}{2m}\epsilon^2\alpha^2(t)[\nabla\theta(\bold{x},\cR(t))]^2,
\label{Masuda2}
\eeqa
where the ``standard potential'' $V_0=V_0({\bold{x}},\cR)$ 
is defined by the stationary GP equation
\beq
\bigg[-\frac{\hbar^2}{2m}\nabla^2+V_0({\bold{x}},\cR)+g\tilde{r}^2({\bold{x}},\cR)\bigg]\tilde{r}({\bold{x}},\cR)={\cal{E}}(\cR)
\tilde{r}(\bold x,\cR).
\label{v0}
\eeq
At the boundary times, but in general only there, $V(\bold x,t)=V_0({\bold{x}},\cR(t))$.   

Equation (\ref{Masuda2}) for the driving potential coincides with 
Eq. (2.28) in \cite{MNProc} for real eigenfunctions, 
whereas Eq. (\ref{Masuda1}) for the phase function $\theta$ 
corresponds to Eq. (2.18) in \cite{MNProc}. 

The present formal framework may be used in the following way:
(i) Starting from a given standard potential $V_0({\bf{x}},\cR)$,
$\tilde{r}(\bold{x},\cR)$ and ${\cal{E}}(\cR)$ would follow from Eq. (\ref{v0}).
Alternatively,  it is also possible to impose $\tilde{r}({\bf{x}},\cR)$ first
and then calculate $V_0$.
(ii) An auxiliary function $\cR(t)$ is imposed.
(iii) $\theta$ has to be determined from Eq. (\ref{Masuda1}).
(iv) The fast-forward potential can be calculated from Eq. (\ref{Masuda2}).

To arrive at this recipe in \cite{MNPra,MNProc} 
preliminary steps are the definition of a standard state, a virtually
fast-forwarded state, and a regularized state with their corresponding equations.
The route followed in Sec. \ref{secmethod} to the driving potential is  
in comparison quite direct. This is so because we made no explicit use of a slow 
reference adiabatic process, 
although it might be deduced from the fast designed dynamics if required. 
The key simplification is to start with the ansatz in Eq. (\ref{wave}) 
and derive the two basic equations for phase and potential from it
by imposing locality and reality of the driving potential. 
Since the phase $\phi$ that solves Eq. (\ref{imag0}) 
depends in general on the particular  $r({\bf{x}},t)$, the potential
calculated through Eqs. (\ref{real}) resp. (\ref{Masuda2}) gives in principle a 
state-dependent potential. However, in some special circumstances,
as we shall see below, the fast-forward potential becomes state independent. 
\section{Connection with invariant's based inverse engineering approach\label{secconexion}}
%
%
%
%
%
%
%
%
In this section we shall relate the previous results for the linear ($g=0$) Schr\"odinger equation to the engineering approach
based on quadratic-in-momentum invariants. 
The non-linear GP equation could also be treated, as in
\cite{GonBEC,BECtransport}, but it 
does not allow in general for
the state-independent potential forms that we shall 
describe for $g=0$.  
\subsection{Lewis-Leach potentials}
%
%
%
In a direct (rather than inverse) approach, the potential $V(\bold{x},t)$ is 
considered to be known, and the wave function at any time $t$ can be deduced 
from the Lewis-Riesenfeld theory of invariants \cite{LR}.  
Suppose that the potential $V(\bold{x},t)$ has the structure of the most general 
``Lewis-Leach'' potential that admits a quadratic-in-momentum invariant \cite{LL},
\beq
\label{LLpot}
V(\bold x,t)=-\bold F(t)\cdot \bold{x}+\frac{1}{2}m\omega^2(t)|\bold{x}|^2+\frac{1}{\rho^2}U({\boldsymbol{\sigma}})+h(t),
\eeq
where $\omega(t)$, $\bold F(t)$ and $h(t)$ are arbitrary functions of time and $U({\boldsymbol{\sigma}})$ is an arbitrary function
of its argument ${\boldsymbol{\sigma}}={\boldsymbol{\sigma}}(t)=(\bold{x}-{\boldsymbol{\alpha}})/\rho$. The time dependent functions $\rho=\rho(t)$ and ${\boldsymbol{\alpha}}={\boldsymbol{\alpha}}(t)$ must satisfy the auxiliary equations
\beqa
\frac{\omega_{0}^{2}}{\rho^{3}}&=&\ddot\rho+\omega^2(t)\rho,
\label{ermakov}\\
\frac{\bold{F}(t)}{m}&=&\ddot{\boldsymbol{\alpha}}+\omega^2(t){\boldsymbol{\alpha}}, \label{osci}
\eeqa
with $\omega_0$ an arbitrary constant. 
The associated dynamical invariant, up to a constant factor, is given by
\beqa
I&=&\frac{1}{2m}|\rho(\bold{p}-m\dot{\boldsymbol{\alpha}})-m\dot\rho(\bold{x}-\boldsymbol{\alpha})|^2 \nonumber\\
&+&\frac{1}{2}m\omega_0^2|\boldsymbol{\sigma}|^2  +U(\boldsymbol{\sigma}), \label{inva0}
\eeqa
with $\bold{p}=-i\hbar\nabla$. 
It satisfies $dI/dt=\partial{I}(t)/\partial{t}
-\frac{i}{\hbar}
[I (t ),H(t )] = 0$, so its expectation values are constant for any wave function $\psi(t)$ that evolves with $H$.

For the potential in Eq. (\ref{LLpot}),
the general solution of the time-dependent Schr\"odinger equation, Eq. (\ref{start}),
can be expanded as a linear combination with constant coefficients $c_n$ and orthonormal eigenvectors $\psi_n$ of $I$ \cite{LR}, 
\beqa
\psi(\bold{x},t)&=&\sum_nc_ne^{i\alpha_n}\psi_n(\bold{x},t), \label{mode} \\
I\psi_n(\bold{x},t)&=&\lambda_n\psi_n(\bold{x},t),
\eeqa
where $\lambda_n$ are the time independent eigenvalues of $I$. The phases $\alpha_n$ satisfy 
$\hbar\frac{d\alpha_n}{dt}=\langle\psi_n|i\hbar\frac{\partial}{\partial t}-H|\psi_n\rangle$, \cite{LR, transport, BECtransport}
\beq
\label{faseLR}
\alpha_n=-\frac{i}{\hbar}\int_{0}^{t}\!\! dt'\bigg(\frac{\lambda_n}{\rho^2}+\frac{m[|\dot{\boldsymbol{\alpha}}\rho-{\boldsymbol{\alpha}}\dot\rho|^2-\omega_{0}^{2}|{\boldsymbol{\alpha}}|^2/\rho^2]}{2\rho^2}+h\bigg). 
\eeq
Performing now the unitary transformation \cite{transport, BECtransport}
\beq
\label{trans}
\psi_n(\bold{x},t)=e^{\frac{im}{\hbar}[\dot\rho |\bold{x}|^2/2\rho+(\dot{\boldsymbol{\alpha}}\rho-{\boldsymbol{\alpha}}\dot\rho)\cdot\bold{x}/\rho]}\frac{1}{\rho^{3/2}}\chi_n({\boldsymbol{\sigma}}),
\eeq
the state $\psi_n$ is easily obtained from the solution $\chi_n(\boldsymbol{\sigma})$ (normalized in $\boldsymbol{\sigma}$-space) of the auxiliary
stationary Schr\"odinger equation
\beq
\bigg[-\frac{\hbar^2}{2m}\nabla^{2}_{{\boldsymbol{\sigma}}}+\frac{1}{2}m\omega_{0}^{2}|{\boldsymbol{\sigma}}|^2+U({\boldsymbol{\sigma}})\bigg]\chi_n({\boldsymbol{\sigma}})=\lambda_n\chi_n({\boldsymbol{\sigma}}).
\label{inva}
\eeq

In the direct approach we assume that 
$U({\boldsymbol{\sigma}})$, $\omega(t)$ and $\bold{F}(t)$ are
known. Solving Eqs. (\ref{ermakov}) and (\ref{osci}) 
we get $\rho(t)$ and 
${\boldsymbol{\alpha}}(t)$ from them. Thus  
Eq. (\ref{inva}) can be solved to get $\lambda_n$ and $\chi_n({\boldsymbol{\sigma}})$.
Finally combining Eqs. (\ref{trans}) and (\ref{inva}), the mode $e^{i\alpha_n}\psi_n$ can be calculated at any time.
\subsection{Inverse engineering approach}
In the inverse approach based on quadratic-in-momentum invariants, the Hamiltonian is assumed to have the form given in Eq. (\ref{LLpot}), at all times and in particular at  
initial and final instants. As $U$ is given the 
stationary Eq. (\ref{inva}) may be solved. Then the  
functions $\rho$ and $\boldsymbol{\alpha}$ are designed so that $[H(t),I(t)]=0$ for $t=0$ and $t=t_f$. Thus  
the Hamiltonian and the invariant have common eigenvectors at these boundary times
\cite{PRL1, transport, BECtransport, 3d, NHSara, OCTexpan, OCTtrans}. 
Typically the initial state $\psi(0)$ is the ground state of $H(0)$ which
is also an eigenstate of $I(0)$, and this state evolves according to Eq. (\ref{mode}), as an eigenvector of the invariant. 

To relate the above to the simple inverse method of Sec. \ref{secmethod} we consider the
single mode wave function $\psi_n$ of Eq. (\ref{trans}) and identify 
\beq
\label{rn}
r_n(\bold x,t)=\chi_n({\boldsymbol{\sigma}})/\rho^{3/2}(t). 
\eeq
The subscript $n$ underlines the dependence with the $n$th mode considered.
Note that $\rho$ and $\boldsymbol{\alpha}$ are chosen at this point.
We may get the phase $\phi_n$ from Eq. (\ref{imag0}).
It can be checked by direct substitution that
\beqa
\phi_n&=&\frac{m}{\hbar}[\dot\rho |\bold{x}|^2/2\rho+(\dot{\boldsymbol{\alpha}}\rho-{\boldsymbol{\alpha}}\dot\rho)\cdot\bold{x}/\rho] \nonumber  \\
&-&\frac{1}{\hbar}\int_{0}^{t}\!\! dt'\frac{\lambda_n}{\rho^2}+\mathcal{F}(t), \label{fn}
\eeqa
where
\beq
\mathcal{F}(t)=-\frac{1}{\hbar}\int_{0}^{t}\!\! dt'\bigg(\frac{m[|\dot{\boldsymbol{\alpha}}\rho-{\boldsymbol{\alpha}}\dot\rho|^2-\omega_{0}^{2}|{\boldsymbol{\alpha}}|^2/\rho^2]}{2\rho^2}+h\bigg),
\eeq
is a solution of this equation. Once $r_n$ and the phase $\phi_n$ are known, 
Eq. (\ref{real}) gives the potential  $V_n(\bold{x},t)$.  
A different arbitrary function of time $\mathcal{F}(t)$ in $\phi_n(\bold{x},t)$ 
would produce a shift of the zero of energy in the resulting potential
$V_n(\bold{x},t)$.
We get, using Eq. (\ref{inva}),
\beqax
V(\bold x,t)&=&-m \left(\ddot{\boldsymbol{\alpha}}+{\boldsymbol{\alpha}}
  \frac{\omega_0^2 + \rho^3 \ddot\rho}{\rho^4}\right) {\boldsymbol{x}}
\\
&+& \frac{m}{2}\left(\frac{\omega_0^2 + \rho^3 \ddot\rho}{\rho^4}\right) |{\boldsymbol{x}}|^2
+ \frac{1}{\rho^2}U({\boldsymbol{\sigma}})+h.
\eeqax
Taking Eqs. (\ref{ermakov}) and (\ref{osci}) into account, this potential
agrees with the potential in Eq. (\ref{LLpot}). 
It is by construction local and real. Moreover it is independent of
the $n$th state considered so that linear combinations  
of the modes at $t=0$ end up at $t_f$ unexcited, preserving the initial 
populations.  

Summarizing, the inverse engineering approach and the fast-forward approach
are connected via the simple inversion method. 
Clearly, the external parameter $\cR$ introduced in Sec. III must be related to ${\boldsymbol{\alpha}}$ and $\rho$, as illustrated in the 
following section.  
%
%
%
%
%
\subsection{Example: harmonic expansion}
Now we discuss an example of a $3D$ harmonic expansion produced with the
inverse engineering approach based on invariants and with the fast forward
technique to illustrate the links between the two methods.

{\it Invariants based approach:} Suppose that the expansion is governed by the Hamiltonian
\beq
\label{Hejem}
H(t)=\frac{\bold{p}^2}{2m}+\frac{1}{2}m\omega^2(t)|\bold{x}|^2,
\eeq
where $\omega(t)$ is unknown, but at the boundary times $\omega(0)=\omega_0$ and $\omega(t_f)=\omega_f$. 
This potential is a particular case of Eq. (\ref{LLpot}) with $\bold{F}(t)=U(\boldsymbol{\sigma})=h(t)=0$. Equation (\ref{osci}) is trivially fulfilled if 
$\boldsymbol{\alpha}(t)=\boldsymbol{\ddot\alpha}(t)=0$ 
and consequently $\boldsymbol{\dot\alpha}(t)=0$.
$\rho(t)$ has to satisfy the Ermakov equation, Eq. (\ref{ermakov}). 
The inverse engineering consists on imposing conditions on $\rho$ and its derivatives,
\beqa
\label{cond}
\rho(0)&=&1,\;\,\quad\dot\rho(0)=0,\;\,\quad\ddot\rho(0)=0,\nonumber\\
\rho(t_f)&=&\gamma,\quad\dot\rho(t_f)=0,\quad\ddot\rho(t_f)=0,
\eeqa
where $\gamma=(\omega_0/\omega_f)^{1/2}$, 
to guarantee the commutation
between $H(t)$ and $I(t)$ at $t=0$ and $t_f$, and then getting $\omega(t)$ from Eq. (\ref{ermakov}). 

{\it Fast forward approach:}
The starting point for the fast forward approach could be the $n$th  eigenstate of
a harmonic trap (see also \cite{MNProc})
with angular frequency ${\cal R}=\omega$,
\beqa
\chi_n(\bold{x},\cR )&=&\frac{\beta^{3/2}e^{-\frac{\beta^2|\bold{x}|^2}{2}}}{\pi^{3/4}\sqrt{2^{n_x+n_y+n_z}n_x!n_y!n_z!}}H_{n_x}(\beta x)
H_{n_y}(\beta y)\nonumber\\
&\times&H_{n_z}(\beta z),
\label{chi}
\eeqa
where $\beta=\sqrt{m\cR/\hbar}$. 
This state plays the role of $\tilde{r}$. 
The corresponding potential $V_0$ is clearly
\beqa
V_0 (x, \cR) &=& \frac{m}{2} R^2 {\bold{x}}^2,
\eeqa
and
\beqa
{\cal{E}}_n&=&\hbar\omega\bigg(n_x+n_y+n_z+\frac{3}{2}\bigg) \label{osciladorE}.
\eeqa
The first step is to solve Eq. (\ref{Masuda1}) and we get as a solution
\beqa
\theta(\bold{x},\cR)=-\frac{m |\bold{x}|^2}{4\hbar\cR}.
\eeqa

{\it Connection:}
The connection between the auxiliary variable $\cR$ in the fast forward
approach and the auxiliary variable $\rho(t)$ in the inverse engineering
approach is in this example explicitely given by
\beq
\cR (t) = \frac{\omega_0}{\rho(t)^2}, 
\eeq
see Eqs. (\ref{rn}) and (\ref{chi}).
The boundary conditions for $\rho(t)$ in Eq. (\ref{cond}) become 
\beqa
\cR(0)&=&\omega_0,\;\,\quad\dot\cR(0)=0,\;\,\quad\ddot\cR(0)=0,\nonumber\\
\cR(t_f)&=&\omega_f,\quad\dot\cR(t_f)=0,\quad\ddot\cR(t_f)=0.
\eeqa
It also follows that $\epsilon\alpha(t)=\dot\cR(t) = -2\omega_0\dot\rho(t)/\rho(t)^3$.
The auxiliary functions $\rho(t)$ resp. $\cR(t)$ can be chosen in some way
fulfilling the boundary conditions.

The corresponding potential in the inverse engineering formalism
is constructed by first solving the Ermakov equation to get $\omega^2(t)$.
Then one has
\beq
V=\frac{1}{2}m\omega^2(t)|\bold{x}|^2=\frac{1}{2}m
\left(\frac{\omega_0^2}{\rho(t)^4} - \frac{\ddot\rho}{\rho}\right)|\bold{x}|^2,
\label{potinv}
\eeq
whereas the fast-forward potential is given, according to Eq. (\ref{Masuda2}),  by
\beqax
\begin{array}{rcccl}
V&=&\frac{m |\bold{x}|^2}{2} &\Big(& \cR^2 + \hbar\epsilon\dot\alpha
\frac{m}{4\hbar\cR} - \hbar\epsilon^2\alpha^2 \frac{m}{4\hbar\cR^2}\\
&&&& - \frac{\hbar^2}{2m}\epsilon^2\alpha^2 \frac{m^2}{4\hbar^2\cR^2}\Big)\\
&=& \frac{m |\bold{x}|^2}{2} &\Big(& \frac{\omega_0^2}{\rho(t)^4} -
\frac{\ddot \rho(t)}{\rho(t)}\Big),
\end{array}
\eeqax
which agrees with Eq. (\ref{potinv}).

\section{Beyond Lewis-Leach potentials: wavefunction splitting processes\label{beyond}}
The transitionless  condition for the inverse engineering method based on invariants relies on the commutativity $[H(t),I(t)]=0$ at times $t=0$ and $t_f$, which 
guarantees common eigenvectors for $H$ and $I$ at these boundary times. 
According to Eq. (\ref{trans}) 
the structure of the density of the $n$th mode of the invariant at initial and final times for quadratic-in-$\bold{p}$ invariants is   
\beq
\label{denI}
\frac{1}{\rho^3 (0)}\bigg|\chi_n\bigg(\frac{\bold x-{\boldsymbol\alpha}(0)}{\rho(0)}\bigg)\bigg|^2\rightarrow\frac{1}{\rho^3 (t_f)}\bigg|\chi_n\bigg(\frac{\bold x-{\boldsymbol\alpha}(t_f)}{\rho(t_f)}\bigg)\bigg|^2.
\eeq
The final density is a translation and/or scaling of the initial one. 
This means that for processes in which the initial and final eigenstates 
of the Hamiltonian do not behave according to Eq. (\ref{denI}), the commutativity 
of $H$ and $I$ at the boundary times cannot be achieved. 
This restriction is due to the use of quadratic-in-$\bold p$ invariants, not to the
invariants-based method. Studying and applying more general invariants is
still an open question. 

As an example in which Eq. (\ref{denI}) does not hold for the final densities, 
let us consider the splitting of an initial state from a single
to a double well potential. For simplicity we take the $1D$ linear Schr\"odinger equation governed by the Hamiltonian
\beq
H(t)=\frac{p^2}{2m}+\frac{1}{2}m\omega^2(t)x^2+\eta(t)x^4.
\eeq
For the initial single trap we consider $\omega^2(0)=\omega_0^2$ and $\eta(0)=\eta_0>0$. The final double well is characterized by a
repulsive harmonic part with $\omega^2(t_f)=-\omega_f^2$ and $\eta(t_f)=\eta_f>0$.
Comparing terms with Eq. (\ref{LLpot}), $F=h=\alpha=0$ and consequently $\dot\alpha=\ddot\alpha=0$, $U(\sigma)=\eta(t)\rho^2x^4$, and  
the particular structure of $U(\sigma)$ sets $\eta(t)=\kappa/\rho^6$, where  
$\rho$ satisfies
the Ermakov equation, Eq. (\ref{ermakov}), and $\kappa$ is an arbitrary constant. The associated invariant is
\beq
I(t)=\frac{1}{2m}(\rho p-m\dot\rho x)^2+\frac{1}{2}m\omega_0^2\frac{x^2}{\rho^2}+\kappa\frac{x^4}{\rho^4}.
\eeq
Imposing $\rho(0)=1$, $\dot\rho(0)=0$, $\ddot\rho(0)=0$, 
and identifying $\kappa=\eta_0$, then $H$ and 
$I$ commute at $t=0$. At $t_f$, $[H(t_f),I(t_f)]=0$ for $\rho(t_f)=(i\omega_0/\omega_f)^{1/2}$, $\dot\rho(t_f)=\ddot\rho(t_f)=0$, and
$\eta_f=-i\eta_0\omega_f^3/\omega_0^3$. 
However, $\rho$ must be a positive real function if initially so, and moreover the final potential that we get is complex. 

\begin{figure}
\begin{center}
\includegraphics[width=9.cm]{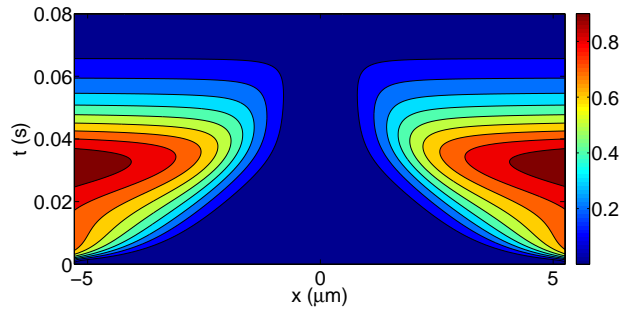}
\caption{(Color online) Phase $\phi(x,t)$ calculated from Eq. (\ref{imag}). Parameters: $m=1.44\cdot 10^{-25}$ kg, $a=3\mu$m, 
$\omega/2\pi=125$ Hz and $t_f=80$ ms.}
\label{fase}
\end{center}
\end{figure}

\begin{figure}
\begin{center}
\includegraphics[width=9.cm]{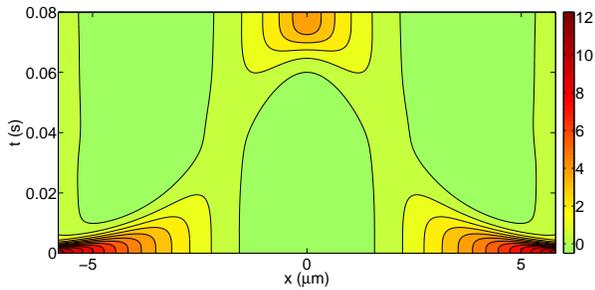}
\caption{(Color online) Fast-forward potential $V(x,t)$ in units of $\hbar\omega$ for a final time $t_f=80$ ms.  The rest of parameters are the same as in Fig. \ref{fase}. Note the coexistence of three wells, 
e.g. around 40 ms, before the final two wells are established.}
\label{pot80}
\end{center}
\end{figure}

\begin{figure}
\begin{center}
\includegraphics[width=9.cm]{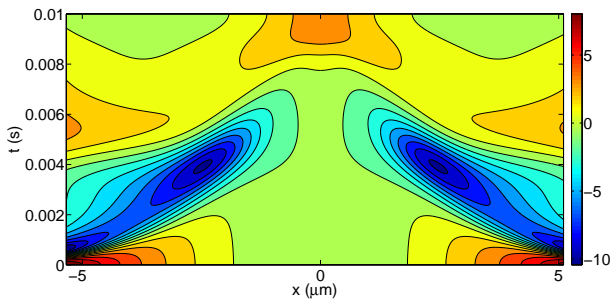}
\caption{(Color online) Fast-forward potential $V(x,t)$ in units of $\hbar\omega$ for a final time $t_f=10$ ms. The rest of parameters are the same as in Fig. \ref{fase}.}
\label{pot10}
\end{center}
\end{figure}

A way out is to use
the simple inverse fast-forward method for specific initial and final states 
without restricting the potential form, see also \cite{MNProc}.
Consider 
for example the
1D splitting 
of 
the initial state $r(x,0)=e^{-\beta^2 x^2/2}$ $(\beta=\sqrt{m\omega/\hbar})$ 
into the final form $r(x,t_f)=e^{-\beta^2 (x-a)^2/2}+e^{-\beta^2(x+a)^2/2}$. 
In between we apply the interpolation
\beq
r(x,t)=z(t)\bigg\{[1-{\cal{R}}(t)]r(x,0)+ 
{\cal{R}}(t)r(x,t_f)\bigg\},
\label{ansatzr}
\eeq
where ${\cal{R}}(t)$ is some smooth, monotonously increasing function
from 0 to 1 and $z(t)$ is a normalization function.
We also impose that $\dot {\cal R}=0$ to ensure $\dot r=0$ at the boundary times $t=0$ and $t_f$.
In the numerical examples the function ${\cal R}(t)$ is chosen as a polynomial of degree 7 to make zero the second and third derivatives at the boundaries. 
Once we have established the form of $r(x,t)$, we solve Eq. (\ref{imag}), ${\rm Im}[V(x,t)]=0$, to get the phase $\phi$ with the
initial conditions $\phi(0,t)=\phi '(0,t)=0$ that fix the zero-energy point (the prime means spatial derivative). Then the resulting phase is introduced into Eq. (\ref{real}) to get the potential $V(x,t)$.   
In Fig. \ref{fase} the phase $\phi(x,t)$ is plotted for a non adiabatic process 
with $t_f=80$ ms. 
The corresponding fast-forward potential is plotted in Fig. \ref{pot80}.
%
%
%
%
%
%
%
%
%
%
$V(x,t)$ in Fig. \ref{pot80} could be realized with high resolution time-varying optical potentials ``painted'' by a tightly focused rapidly moving laser 
beam \cite{HRM09}, or by means of spatial light modulators \cite{Foot}.
A simpler approximate approach would involve the 
combination of three Gaussian beams. 
In principle the time $t_f$ can be reduced to produce the splitting in a shorter time. For example, in Fig. \ref{pot10} $t_f=10$ ms and a more complicated potential is needed. 
%
%
%
%
%
%
%
%
%
%

%
%
%
%
%
%
%
%
\section{Discussion\label{secoutlook}}
We have first distilled from the somewhat imposing set of equations 
of the fast-forward (FF) formalism as originally presented a streamlined version that 
may aid to apply it more easily. 
Our second aim has been to relate it to other inverse engineering methods.    
In a previous publication, the inverse-engineering method based on invariants,  was related to the transitionless tracking algorithm, and their potential equivalence was demonstrated \cite{CTM}. 
Similarly we have established in this paper the connection between the fast-forward method and the invariant-based  method for quadratic-in-momentum invariants. 
These relations do not imply the full identity of  the methods but their overlap and equivalence in a common domain. They are still useful heuristically as separate approaches since they are formulated in rather different terms \cite{CTM,MN11}. Moreover they  facilitate extensions beyond their common domain, as exemplified
by the wave-splitting processes discussed in the previous section.    
Further extensions are left for separate analysis: for example the possibility to transfer an excited state into the ground state or viceversa, or combining the fast-forward approach with optimal control theory (OCT) without including the final fidelity in the cost function as in \cite{S07, S09a, S09b}. (This would be possible because the fidelity is guaranteed to be one by construction.)
It will also be interesting for future work to consider complex potentials, either as solutions to the shortcut dynamics, as in the quantum brachistochrone \cite{qb}, or as 
an effective description of the system dynamics to be accelerated \cite{NHSara}.          

We are grateful to S. Masuda and K. Nakamura for discussing their method. 
We acknowledge funding by Projects No. GIU07/40 and No. FIS2009-12773-C02-01, 
and the UPV/EHU
under program UFI 11/55.  
E. T. acknowledges financial support from the Basque Government 
(Grants No. BFI08.151).

\end{document}